\documentclass[aps,prl,groupedaddress,fleqn,twocolumn,10pt]{revtex4}
\pdfoutput=1
\usepackage{graphicx}
\usepackage{amsmath}
\usepackage{gensymb}
\usepackage{float}
\usepackage{physics}
\usepackage{xcolor}

\begin{document}

\raggedbottom

\title{Speed of sound from fundamental physical constants}
\author{K. Trachenko}
\affiliation{School of Physics and Astronomy, Queen Mary University of London, Mile End Road, London, E1 4NS, UK}
\author{B. Monserrat}
 \affiliation{Department of Materials Science and Metallurgy, University of Cambridge, 27 Charles Babbage Road, Cambridge CB3 0FS, United Kingdom}%
 \affiliation{Cavendish Laboratory, University of Cambridge, J. J. Thomson Avenue, Cambridge CB3 0HE, United Kingdom}
\author{C. J. Pickard}
 \affiliation{Department of Materials Science and Metallurgy, University of Cambridge, 27 Charles Babbage Road, Cambridge CB3 0FS, United Kingdom}%
\affiliation{Advanced Institute for Materials Research, Tohoku University, Sendai, Japan}
\author{V. V. Brazhkin}
\affiliation{Institute for High Pressure Physics, RAS, 108840, Troitsk, Moscow, Russia}

\begin{abstract}
Two dimensionless fundamental physical constants, the fine structure constant $\alpha$ and the proton-to-electron mass ratio $\frac{m_p}{m_e}$ are attributed a particular importance from the point of view of nuclear synthesis, formation of heavy elements, planets, and life-supporting structures. Here, we show that a combination of these two constants results in a new dimensionless constant which provides the upper bound for the speed of sound in condensed phases, $v_u$. We find that $\frac{v_u}{c}=\alpha\left(\frac{m_e}{2m_p}\right)^{\frac{1}{2}}$, where $c$ is the speed of light in vacuum. We support this result by a large set of experimental data and first principles computations for atomic hydrogen. Our result expands current understanding of how fundamental constants can impose new bounds on important physical properties.
\end{abstract}

\maketitle

\section{Introduction}

Several notable properties of condensed matter phases are defined by fundamental physical constants. The Bohr radius gives a characteristic scale of interatomic distance on the order of the Angstrom, in terms of electron mass $m_e$, charge $e$, and Planck constant $\hbar$. These same fundamental constants enter the Rydberg energy, setting the scale of a characteristic bonding energy in condensed phases and chemical compounds \cite{ashcroft}.

Among the fundamental constants, those that are {\it dimensionless} and do not depend on the choice of units, play a special role in physics \cite{barrow}. Two important dimensionless constants are the fine structure constant $\alpha$ and the proton-to-electron mass ratio, $\frac{m_p}{m_e}$. The finely-tuned values of $\alpha$ and $\frac{m_p}{m_e}$, and the balance between them, governs nuclear reactions such as proton decay and nuclear synthesis in stars, leading to the creation of the essential biochemical elements, including carbon. This balance provides a narrow ``habitable zone'' in the ($\alpha$,$\frac{m_p}{m_e}$) space where stars and planets can form and life-supporting molecular structures can emerge \cite{barrow}.

We show that a simple combination of $\alpha$ and $\frac{m_p}{m_e}$ results in another dimensionless quantity which has an unexpected and specific implication for a key property of condensed phases, the speed at which waves travel in solids and liquids, or the speed of sound, $v$. We find that this combination provides an upper bound for $v$, $v_u$, as
\begin{equation}
\frac{v_u}{c}=\alpha\left(\frac{m_e}{2m_p}\right)^{\frac{1}{2}},
\label{v0}
\end{equation}
\noindent where $c$ is the speed of light in vacuum.

We support this result with a large set of experimental data for different systems, and the first principles modelling of atomic hydrogen.

Identifying and understanding bounds on physical properties is important from the point of view of fundamental physics, predictions for theory and experiment, as well as searching for and rationalizing universal behavior (see, e.g., \cite{kss,zaanen3,hartnoll,zaanen2,spin,behnia,zaanen1,behnia1,hartnoll1}). Properties for which bounds were recently discussed include viscosity and diffusivity. The proposed {\it lower} bounds for these two properties feature in a range of areas including, for example, strongly-interacting field theories, quark-gluon plasmas, holographic duality, electron diffusion, transport properties in metals and superconductors, and spin transport in Fermi gases \cite{kss,zaanen3,hartnoll,zaanen2,spin,behnia,zaanen1,behnia1,hartnoll1}. Recently, two of us found a lower bound for the kinematic viscosity of liquids set by fundamental physical constants \cite{sciadv}. Here, we propose a new, {\it upper}, bound for the speed of sound in condensed matter phases in terms of fundamental constants.

Apart from setting the speed of elastic interactions in solids, $v$ is related to elasticity, hardness and affects important low-temperature thermodynamic properties such as energy, entropy and heat capacity \cite{landau}. As discussed below, the upper bound of $v$ sets the smallest possible entropy and heat capacity at a given temperature.

In solids, $v$ depends on elastic properties and density. These strongly depend on the bonding type and structure which are inter-dependent \cite{phillips}. As a result, it was not thought that $v$ can be predicted analytically without simulations, contrary to other properties such as energy or heat capacity which are universal in the classical harmonic approximation \cite{landau}. In view of this, representing the upper bound of $v$ in terms of fundamental constants is notable.

\section{Results and discussion}

There are two approaches in which $v$ can be evaluated. The two approaches start with system elasticity and vibrational properties, respectively.

We begin with system elasticity. The longitudinal speed of sound is $v=\left({\frac{M}{\rho}}\right)^{\frac{1}{2}}$, where $M=K+\frac{4}{3}G$, $K$ is the bulk modulus, $G$ is the shear modulus, and $\rho$ is the density. It has been ascertained that elastic constants are governed by the density of electromagnetic energy in condensed matter phases. In particular, a clear relation was established between the bulk modulus $K$ and the bonding energy $E$: $K=f\frac{E}{a^3}$, where $a$ is the interatomic separation and $f$ is the proportionality coefficient \cite{diamond,diamond1}. This relation can be derived up to a constant given by the second derivative of the function representing the dependence of energy on volume. For a majority of strongly-bonded solids, $f$ varies in the range 1-4 \cite{diamond,diamond1}. The same data implies the proportionality coefficient between $M$ and $\frac{E}{a^3}$ in the range of about 1-6. Combining $v=\left({\frac{M}{\rho}}\right)^{\frac{1}{2}}$ and $M=f\frac{E}{a^3}$ gives $v=f^{\frac{1}{2}}\left(\frac{E}{m}\right)^{\frac{1}{2}}$, where $m$ is the mass of the atom or molecule, and we used $m=\rho a^3$. The factor $f^{\frac{1}{2}}$ is about 1-2 and can be dropped in an approximate evaluation of $v$. Then,

\begin{equation}
v=\left(\frac{E}{m}\right)^{\frac{1}{2}}.
\label{v01}
\end{equation}

We now recall that the bonding energy in condensed phases is given by the Rydberg energy on the order of several eV \cite{ashcroft} as
\begin{equation}
E_{\rm R}=\frac{m_ee^4}{32\pi^2\epsilon_0^2\hbar^2},
\label{rydberg}
\end{equation}
\noindent where $e$ and $m_e$ are electron charge and mass.

$E_{\rm R}$ is used for order-of-magnitude estimations of the bonding energy $E$ \cite{ashcroft}. Using $E=E_{\rm R}$ from (\ref{rydberg}) in (\ref{v01}) gives
\begin{equation}
v=\alpha\left(\frac{m_e}{2m}\right)^{\frac{1}{2}}c,
\label{v00}
\end{equation}
\noindent where $\alpha=\frac{1}{4\pi\epsilon_0}\frac{e^2}{\hbar c}$ is the fine structure constant.

A result similar to (\ref{v00}) can be obtained in the second approach that starts with the consideration of the vibrational properties of the system. The longitudinal speed of sound, $v$, can be evaluated as the phase velocity from the longitudinal dispersion curve $\omega=\omega$($k$) in the Debye approximation: $v=\frac{\omega_{\rm D}}{k_{\rm D}}$, where $\omega_{\rm D}$ and $k_{\rm D}$ are Debye frequency and wavevector, respectively. Using $k_{\rm D}=\frac{\pi}{a}$, where $a$ is the interatomic (inter-molecule) separation, gives
\begin{equation}
v=\frac{1}{\pi}\omega_{\rm D}a.
\label{v001}
\end{equation}

We recall that the characteristic scale of interatomic separation is given by the Bohr radius $a_{\rm B}$ on the order of the Angstrom as
\begin{equation}
a_{\rm B}=\frac{4\pi\epsilon_0\hbar^2}{m_e e^2}.
\label{bohr}
\end{equation}

We now use the known ratio between the phonon energy, $\hbar\omega_{\rm D}$, and $E$. The phonon energy $\hbar\omega_{\rm D}$ can be approximated as $\hbar\left(\frac{E}{ma^2}\right)^{\frac{1}{2}}$, where $m$ is the mass of the atom. Taking the ratio $\frac{\hbar\omega_{\rm D}}{E}$ and using $a=a_{\rm B}$ from (\ref{bohr}) and $E=E_{\rm R}$ from (\ref{rydberg}) gives $\frac{\hbar\omega_{\rm D}}{E}$, up to a constant factor close to unity, as
\begin{equation}
\frac{\hbar\omega_{\rm D}}{E}=\left(\frac{m_e}{m}\right)^{\frac{1}{2}}.
\label{ratio}
\end{equation}

Using (\ref{ratio}) in (\ref{v001}) gives
\begin{equation}
v=\frac{Ea}{\pi\hbar}\left(\frac{m_e}{m}\right)^{\frac{1}{2}}.
\label{v1}
\end{equation}

$v$ in (\ref{v00}), up to a constant factor, can now be obtained by using $a=a_{\rm B}$ from (\ref{bohr}) and $E=E_{\rm R}$ from (\ref{rydberg}) in (\ref{v1}). Alternatively, the same result can be found by (a) recalling that the bonding energy, or the characteristic energy of electromagnetic interaction, is $E=\frac{\hbar^2}{2m_ea^2}$ and (b) using this $E$ and $a=a_{\rm B}$ (\ref{bohr}) in (\ref{v1}).

As compared to the first approach, the second approach to evaluating $v$ involves additional approximations, including evaluating $v$ from the dispersion relation in the Debye model, using $a=a_{\rm B}$ in (\ref{bohr}), and the ratio between the phonon and bonding energies (\ref{ratio}). We therefore focus on the result from the first approach, Eq. (\ref{v00}).

We now discuss Eq. (\ref{v00}) and its implications. $m_e$ characterises electrons, which are responsible for the interactions between atoms. The electronic contribution is further reflected in the factor $\alpha c$ ($\alpha c\propto\frac{e^2}{\hbar}$), which is the electron velocity in the Bohr model.

We note that $\alpha c$ and $v$ do not depend on $c$. The reason for writing $v$ in terms of $\alpha c$ in Eq. (\ref{v00}) and the ratio $\frac{v_u}{c}$ in terms of $\alpha$ in Eq. (\ref{v0}) is two-fold. First, it is convenient and informative to represent the bound in terms of the ratio $\frac{v_u}{c}$, similarly to the ratio of the Fermi velocity and the speed of light $\frac{v_{\rm F}}{c}$ commonly used. Second, it is $\alpha$ (together with $\frac{m_p}{m_e}$) that is given fundamental importance and is finely tuned to result in proton stability and to enable the synthesis of heavy elements \cite{barrow} and, therefore, the existence of solids and liquids where sound can propagate to begin with.

$m$ in (\ref{v00}) characterises atoms involved in sound propagation. Its scale is set by the proton mass $m_p$: $m=Am_p$, where $A$ is the atomic mass. Recall that $a_{\rm B}$ in (\ref{bohr}) and $E_{\rm R}$ in (\ref{rydberg}) are characteristic values derived for the H atom. We similarly set $A=1$ and $m=m_p$ in (\ref{v00}) to arrive at the upper bound of $v$ in (\ref{v00}), $v_u$, as

\begin{equation}
v_u=\alpha\left(\frac{m_e}{2m_p}\right)^{\frac{1}{2}}c~\approx~36,100~\frac{\rm m}{\rm s},
\label{v3}
\end{equation}

\noindent and observe that $v_u$ depends on fundamental physical constants only, including the dimensionless fine structure constant $\alpha$ and the proton-to-electron mass ratio.

Equation (\ref{v3}) is the extension of (\ref{v00}) to atomic hydrogen. We will calculate $v$ in atomic H later in the paper.

Combining Eqs. (\ref{v00}), (\ref{v3}), and $m=Am_p$ gives
\begin{equation}
v=\frac{v_u}{A^\frac{1}{2}}.
\label{a}
\end{equation}

Before discussing the experimental data in relation to Eq. (\ref{v00}) and its consequences, Eqs. (\ref{v3})-(\ref{a}), we note that the speed of sound is governed by the elastic moduli and density which substantially vary with bonding type: from strong covalent, ionic, or metallic bonding, typically giving a large bonding energy to intermediate hydrogen-bonding, and weak dipole and van der Waals interactions. Elastic moduli and density also vary with the particular structure that a system adopts. Furthermore, the bonding type and structure are themselves inter-dependent: covalent and ionic bonding result in open and close-packed structures, respectively \cite{phillips}. As a result, the speed of sound for a particular system can not be predicted analytically and without the explicit knowledge of structure and interactions \cite{zaccone}, similarly to other system-dependent properties such as viscosity or thermal conductivity (but differently to other properties such as the classical energy and specific heat which are universal in the harmonic approximation \cite{landau}). Nevertheless, the dependence of $v$ on $m$ or $A$ can be studied in a family of elemental solids. Elemental solids do not have confounding features existing in compounds due to mixed bonding between different atomic species (including mixed covalent-ionic bonding between the same atomic pairs as well as different bonding types between different pairs).

\begin{figure}
{\scalebox{0.37}{\includegraphics{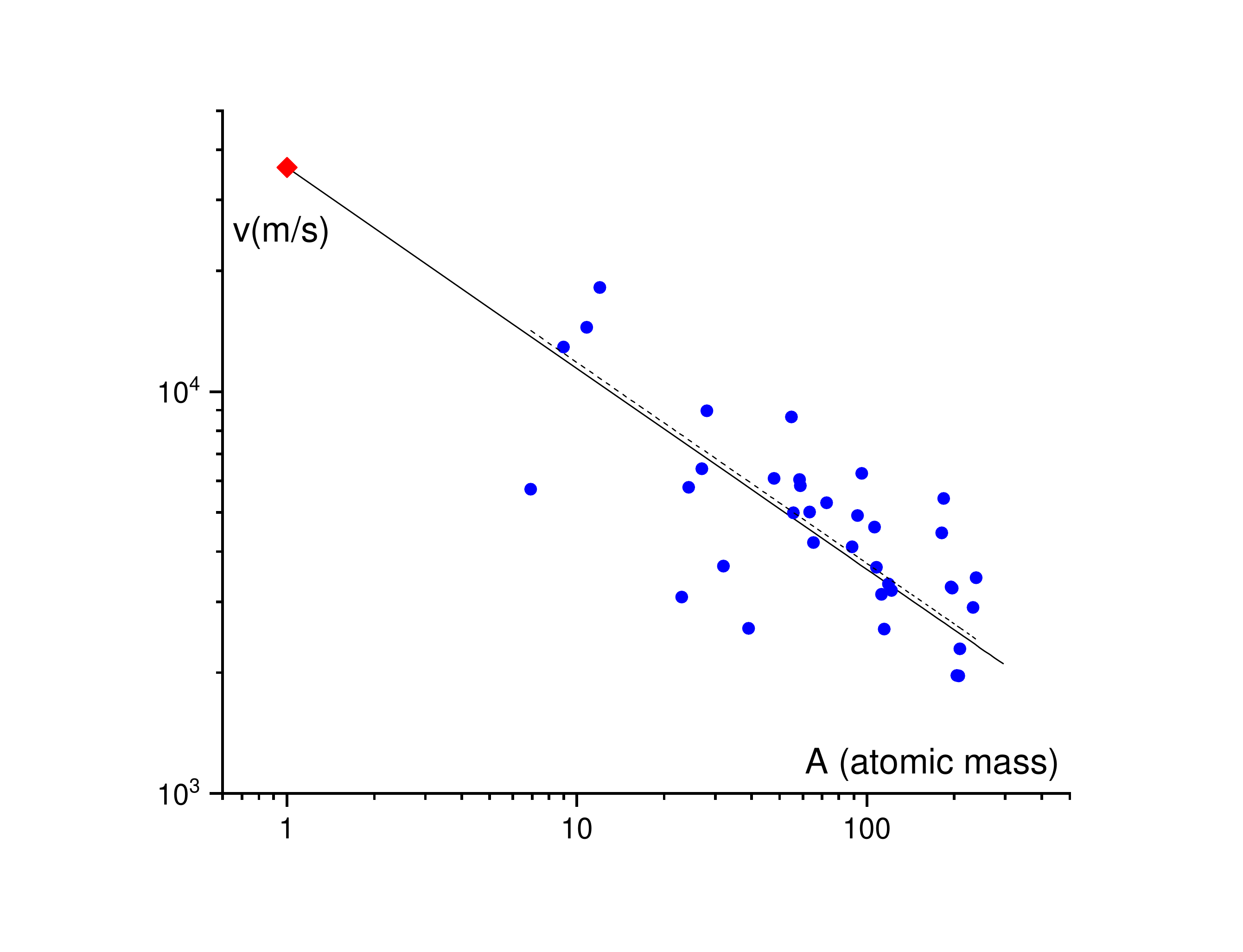}}}
\caption{Experimental longitudinal speed of sound \cite{handbook,handbook1,handbook2} in 36 elemental solids (blue bullets) as a function of atomic mass. The solid line is the plot of Eq. (\ref{a}): $v=\frac{v_u}{A^\frac{1}{2}}$. The red diamond shows the upper bound of the speed of sound (\ref{v3}). The dashed line is the fit to the experimental data points. In order of increasing mass, the solids are: Li, Be, B, C, Na, Mg, Al, Si, S, K, Ti, Mn, Fe, Ni, Co, Cu, Zn, Ge, Y, Nb, Mo, Pd, Ag, Cd, In, Sn, Sb, Ta, W, Pt, Au, Tl, Pb, Bi, Th and U.
}
\label{elemental}
\end{figure}

To compare Eq. (\ref{a}) to experiments, we plot the available data of $v$ as a function of $A$ for 36 elemental solids \cite{handbook,handbook1,handbook2} in Fig. \ref{elemental}, including semiconductors and metals with large bonding energies. The data are depicted in a log-log plot. Equation (\ref{a}) is the straight line in Fig. \ref{elemental} ending in its upper theoretical bound (\ref{v3}) for $A=1$. The linear Pearson correlation coefficient calculated for the experimental set ($\log$ A, $\log$ v) is $-0.71$. Its absolute value is slightly above the boundary notionally separating moderate and strong correlations \cite{correlation}. The ratio of calculated and experimental $v$ is in the range 0.6-2.4, consistent with the range of $f^\frac{1}{2}$ approximated by 1 in the derivation of Eq. (\ref{v01}).

We also show the fit of the experimental data points to the inverse square root function predicted by Eq. (\ref{a}) as the dashed line in Fig. \ref{elemental} and observe that it lies close to Eq. (\ref{a}). The fitted curve gives the intercept at 37,350 $\frac{\rm m}{\rm s}$, in about 3\% agreement with $v_u$ in (\ref{v3}). This indicates that the numerical coefficient in Eq. (\ref{v00}), which is subject to an approximation as mentioned earlier, and discussed below in more detail, gives good agreement with the experimental trend.

\begin{figure}
{\scalebox{0.37}{\includegraphics{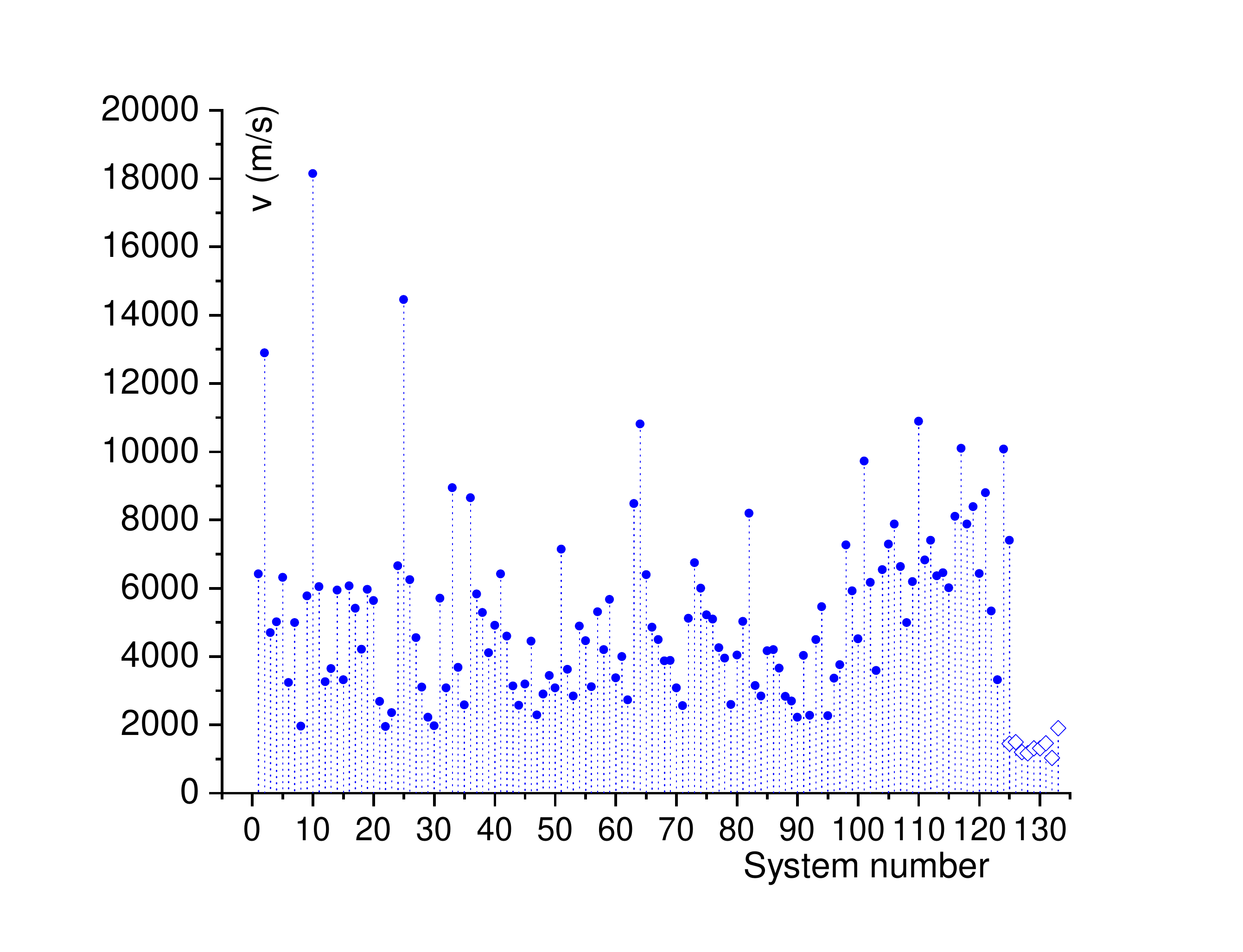}}}
\caption{Experimental longitudinal speed of sound \cite{handbook,handbook1,handbook2} in 124 solids (circles) and 9 liquids \cite{handbook} (diamonds) at ambient conditions as a function of the system number. Solids are: Al, Be, Brass, Cu, Duralumin, Au, Fe, Pb, Mg, Diamond, Ni, Pt, Ag, Steel, Sn, Ti, W, Zn, Fused silica, Pyrex glass, Lucite, Polyethylene, Polyesterene, WC, B, Mo, NaCl, RbCl, RbI, Tl, Li, Na, Si, S, K, Mn, Co, Ge, Y, Nb, Mo, Pd, Cd, In, Sb, Ta, Bi, Th, U, LiF, LiCl, BeO, NH$_4$H$_2$PO$_4$, NH$_4$Cl, NH$_4$Br, NaNO$_3$, NaClO$_3$, NaF, NaBr, NaBrO$_3$, NaI, Mg$_2$SiO$_4$, $\alpha$-Al$_2$O$_8$, AlPO$_4$, AlSb, KH$_2$PO$_4$, KAl(SO$_4$)$_2$, KCl, KBr, KI, CaBaTiO$_3$, CaF$_2$, ZnO, $\alpha$-ZnS, GaAs, GaSb, RbF, RbBr, Sr(NO$_3$)$_2$, SrSO$_4$, SrTiO$_3$, AgCl, AgBr, CdS, InSb, CsCl, CsBr, CsI, CsF, Ba(NO$_3$)$_2$, BaF$_2$, BaSO$_4$, BaTiO$_3$, TlCl, Pb(NO$_3$)$_2$, PbS, Apatite, Aragonite, Barite, Beryl, Biotite, Galena, Hematite, Garnet, Diopside, Calcite, Cancrinite, Alpha-quartz, Corundum, Labradorite, Magnetite, Microcline, Muscovite, Nepheline, Pyrite, Rutile, Staurolite, Tourmaline, Phlogopite, Chromite, Celestine, Zircon, Spinel and Aegirite. Liquids are: Mercury, Water, Acetone, Ethanol, Ethylene, Benzene, Nitrobenzene, Butane and Glycerol. See Refs. \cite{handbook,handbook1,handbook2} for system specifications, including density and symmetry groups.
}
\label{all}
\end{figure}

The agreement of Eq. (\ref{a}) with experimental data supports Eq. (\ref{v00}) and its consequence, the upper limit $v_u$ in Eq. (\ref{v3}). We now show that $v_u$ agrees with a wider experimental set. In Fig. \ref{all}, we show experimental $v$ \cite{handbook,handbook1,handbook2} in 133 systems, including compounds together with the elemental solids in Fig. \ref{elemental}. We observe that experimental $v$ are smaller than the upper theoretical bound $v_u$ in (\ref{v3}). $v_u$ is about twice as large as $v$ in diamond, the highest speed of sound measured at ambient conditions (the in-plane speed of sound in graphite is slightly above $v$ in diamond \cite{behnia1}).

Eq. (\ref{a}) can be used to roughly predict the average, or characteristic speed of sound $v$. $A^{\frac{1}{2}}$ which, according to (\ref{a}) is relevant for the speed of sound, varies across the periodic table in the range of about 1-15, with an average value of 8. According to (\ref{a}), the corresponding $v$ is $v\approx 4,513\frac{\rm m}{\rm s}$. This is in 16\% agreement with 5,392 $\frac{\rm m}{\rm s}$, the average over all elemental solids and in 14\% agreement with 5,267 $\frac{\rm m}{\rm s}$, the average over all solids in Fig. \ref{all}.

We have included the experimental speed of sound of room-temperature liquids in Fig. \ref{all}, with typical $v$ in the range 1,000-2,000 $\frac{\rm m}{\rm s}$. $v$ in high-temperature liquid metals such as Al, Fe, Mg, and Ni extends to higher values in the range 4,000-5,000 $\frac{\rm m}{\rm s}$ \cite{metals}. Similarly to solids, $v$ in liquids satisfy the bound $v_u$. We note that our evaluation of $v$ and $v_u$ applies to liquids with cohesive states \cite{f2}, where molecular dynamics includes solid-like oscillatory components \cite{frenkel} and where $v$ is set by the elastic moduli as in solids, albeit taken at their high-frequency (short-time) values \cite{frenkel,boon}. On the other hand, at high temperature and/or low density, cohesive states are lost and Eq. (\ref{rydberg}) and Eq. (\ref{bohr}) and our derivation of $v$ do not apply. In this regime, the moduli are related to the kinetic energy of molecules rather than interactions and bonding energy, and $v$ starts to increase with temperature and loses its universality. Above the Frenkel line \cite{f1,f2,f3}, formalising the qualitative change of molecular dynamics from combined oscillatory and diffusive to purely diffusive, $v$ is equal to the thermal speed of molecules as in a gas.

With regard to liquids, we note that an expression similar to (\ref{v01}) was earlier obtained by evaluating the elastic modulus using the liquid state theory and applied to liquid metals \cite{gitis}. The speed of sound can also be evaluated in the theory of metals using the ionic plasma frequency and subsequently accounting for the conduction electrons screening. This results in the Bohm-Staver relation $v\propto\left(\frac{m_e}{m}\right)^{\frac{1}{2}}v_{\rm F}$, where $v_{\rm F}$ is the Fermi velocity \cite{ashcroft}, and hence $v\propto\frac{1}{A^\frac{1}{2}}$ as in Eq. (\ref{a}) (the factor $\left(\frac{m_e}{m}\right)^{\frac{1}{2}}$ also appears in the ratio of sound to melting velocity \cite{hartnoll1}). These and other relations derived for the liquid state give a fairly good account of the experimental sound velocity in liquid metals \cite{gitis,metals}.

We make three further remarks about the calculated $v$ and its bound. First, this derivation involves approximations as mentioned earlier. The approximations may affect the numerical factor in Eqs. (\ref{v00}) and (\ref{v3}). At the same time, the characteristic scale of $v$ in (\ref{v00}) and its upper bound (\ref{v3}) is set by fundamental constants. Second, Eq. (\ref{rydberg}) as well as Eqs. (\ref{bohr})-(\ref{ratio}) used in the second derivation of $v$ assume valence electrons directly involved in bonding and hence strongly-bonded systems, including metallic, covalent and ionic solids. Although bonding in weakly-bonded solids such as noble, molecular and hydrogen-bonded solids is also electromagnetic in origin, weak dipole and van der Waals interactions result in smaller $E$ \cite{vadim1} and smaller $v$ as a result. Therefore, the upper bound $v_u$ applies to weakly-bonded systems too. We note here that our evaluation does not directly distinguish between bonding types and hence does not consider the trend of $v$ to increase along the rows of the periodic table, from soft metals to hard covalent materials in Fig. \ref{elemental}. This trend can be accounted for by (a) noting that $v$ in (\ref{v1}) and $E=\frac{\hbar^2}{2m_ea^2}$ imply $v\propto\frac{1}{a}$ and (b) introducing an extra parameter into the equation for $v$ related to density (we are grateful to K. Behnia for pointing this out). Third, our evaluation of $v$ does not account for the effect of pressure on $E$ and $a$ and applies when the enthalpic term is relatively small.


Our upper bound in Eq. (\ref{v3}) corresponds to solid hydrogen with strong metallic bonding. Although this phase only exists at megabar pressures \cite{silvera,loubeyre} and is dynamically unstable at ambient pressure where molecular formation occurs, it is interesting to calculate $v$ in atomic hydrogen in order to check the validity of our upper bound. In addition, there has been strong interest in the properties of atomic hydrogen at high pressure (see, e.g., Refs. \cite{silvera,loubeyre,hydrogen}), although the speed of sound in these phases was not discussed and remains unknown.

\begin{figure}
{\scalebox{0.33}{\includegraphics{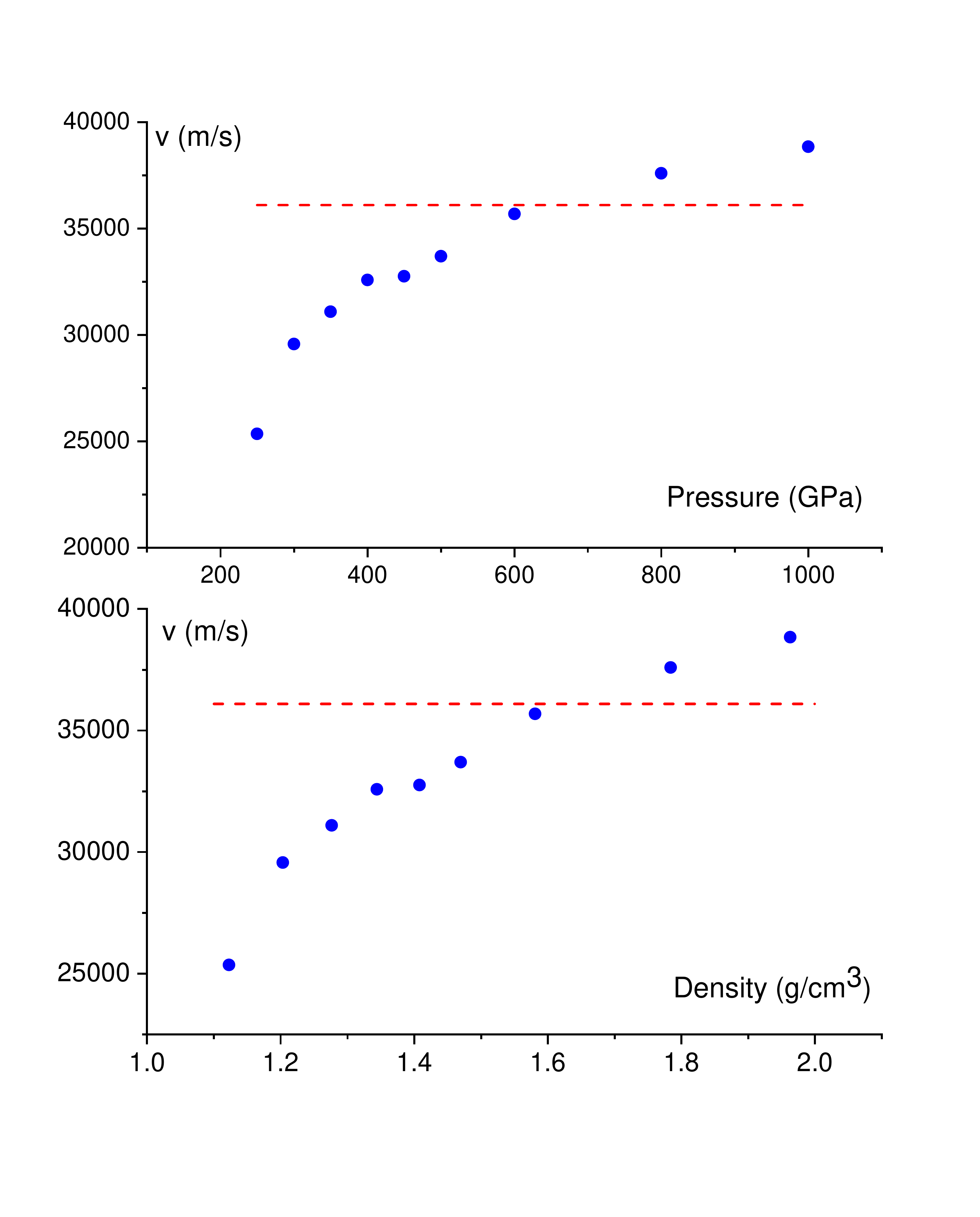}}}
\caption{Calculated speed of sound in atomic hydrogen as a function of pressure (top) and density (bottom). The dashed line shows the upper bound $v_u$ in (\ref{v3}).}
\label{abinitio}
\end{figure}

We have calculated the speed of sound in atomic hydrogen for the $I4_1/amd$ structure \cite{i41amd,pickard_h_natphys}, which is currently the best candidate structure for solid atomic metallic hydrogen. This structure is calculated to become thermodynamically stable in the pressure range $400$--$500$ GPa \cite{azadi_metal,morales_metal}, below which solid hydrogen is a molecular solid. However, we find that $I4_1/amd$ is dynamically stable at pressures above about $250$ GPa, and therefore we perform calculations in the pressure range $250$--$1000$ GPa. The speed of sound as a function of pressure and density reported in Fig. \ref{abinitio} corresponds to the highest energy acoustic branch and is averaged over stochastically generated directions in $\mathbf{q}$-space.

Our upper bound (\ref{v3}) does not account for the enthalpic contribution to the system energy as mentioned earlier; including the pressure effect would increase $v_u$ considerably at pressures in Fig. \ref{abinitio}. Despite this, the calculated $v$ remains below $v_u$ in a wide pressure range and starts increasing above $v_u$ only above very high pressure of about 600 GPa. In this regard, we note that hydrogen is a unique element with no core electrons. This results in the absence of strong repulsive contributions to the interatomic interaction as compared with heavier elements and, consequently, weaker pressure dependence of elastic moduli and the speed of sound \cite{vadim2}. We also note that sharper change of $v$ at lower pressure in Fig. \ref{abinitio} is related to approaching the limit of dynamical stability of the $I4_1/amd$ structure around 250 GPa.

We make three remarks related to previous work. It was noted that thermal diffusivity of insulators does not fall below a threshold value given by the product of $v^2$ and the Planckian time \cite{behnia}. Later work linked the upper bound on the speed of sound to the melting velocity related to melting temperature and Lindemann criterion \cite{hartnoll1}. Finally, the upper bound of the speed of sound for hadronic matter was conjectured as $\frac{c}{\sqrt{3}}$ and discussed (see, e.g., Ref. \cite{hadronic}) for review). Comparing this bound with (\ref{v0}), we see that our bound is smaller due to small coupling constant $\alpha$ and the electron-to-proton mass ratio. In hadronic matter with strong coupling and particles with the same or similar masses, these factors become on the order of 1, in which case our $\frac{v_u}{c}$ in Eq. (\ref{v0}) becomes closer to the conjectured limit \cite{hadronic}.

As discussed above, $v$ features in several thermodynamic properties of solids. For example, the low-temperature entropy and heat capacity per volume are $\frac{S}{V}=\frac{2\pi^2}{15(\hbar u)^3}T^3$ and $\frac{C}{V}=\frac{2\pi^2}{5(\hbar u)^3}T^3$, where $u$ is the average speed of sound and $k_{\rm B}=1$ \cite{landau}. Hence, the upper bound for $u$ gives the smallest possible entropy and heat capacity at a given temperature.

\section{Materials and Methods}

We have performed density functional theory calculations using the {\sc castep} package \cite{castep}, with the Perdew-Burke-Ernzerhof (PBE) exchange-correlation functional \cite{pbe}, an energy cutoff of $1200$\,eV and a $\mathbf{k}$-point grid of spacing $2\pi\times0.025$\,\AA$^{-1}$ to sample the electronic Brillouin zone. We have relaxed the cell parameters and internal coordinates to obtain a pressure to within $10^{-4}$ GPa of the target pressure and forces smaller than $10^{-5}$\,eV/\AA. We have then calculated the phonon spectrum using the finite difference method \cite{fd_martin} in conjunction with nondiagonal supercells \cite{nondiagonal} with a $4\times4\times4$ coarse $\mathbf{q}$-point grid to sample the vibrational Brillouin zone. We have used Fourier interpolation to calculate the phonon frequencies at $\mathbf{q}$-vectors close to the $\Gamma$-point and then used finite differences to calculate the corresponding speed of sound.

\section{Conclusions}

We conclude by returning to dimensionless fundamental physical constants. Rewriting (\ref{v3}) as
\begin{equation}
\frac{v_u}{c}=\alpha\left(\frac{m_e}{2m_p}\right)^{\frac{1}{2}},
\label{v4}
\end{equation}
\noindent we observe that the combination of two important dimensionless fundamental constants, the fine structure constant $\alpha$ and the electron-to-proton mass ratio, interestingly gives the new dimensionless ratio, $\frac{v_u}c$.

\section{Acknowledgements}

We are grateful to M. Baggioli, K. Behnia, S. Hartnoll, J. Zaanen and A. Zaccone for discussions. C.J.P. is supported by the Royal Society through a Royal Society Wolfson Research Merit Award and the EPSRC through Grant No. EP/P022596/1. K. T. acknowledges the EPSRC support.

Funding: We are grateful to the EPSRC for support.

Author contributions: the authors have contributed equally to this paper.

Competing interests: The authors declare no competing interests.

Data and materials availability: All data needed to evaluate the conclusions in the paper are present in the paper and references. Additional data related to this paper may be requested from the authors.


\begin{thebibliography}{99}

\bibitem{ashcroft} N. W. Ashcroft and N. D. Mermin, {\it Solid State Physics} (Saunders College Publishing, 1976).

\bibitem{barrow} J. D. Barrow, {\it The Constants of Nature} (Pantheon Books, 2003).

\bibitem{kss} P. K. Kovtun, D. T. Son and A. O. Starinets, Starinets, Viscosity in strongly interacting quantum field
theories from black hole physics. {\it Phys. Rev. Lett.} {\bf 94}, 111601 (2005).

\bibitem{zaanen3} J. Zaanen, Why the temperature is high. {\it Nature} {\bf 430}, 512-513 (2004).

\bibitem{hartnoll} S. A. Hartnoll, Theory of universal incoherent metallic transport. {\it Nat. Phys.} {\bf 11}, 54-61 (2015).

\bibitem{zaanen2} J. Zaanen, Planckian dissipation, minimal viscosity and the transport in cuprate strange
metals. {\it SciPost Phys.} {\bf 6}, 061 (2019).

\bibitem{spin} C. Luciuk et al, Observation of quantum-limited spin transport in strongly interacting two-dimensional
Fermi gases. {\it Phys. Rev. Lett.} {\bf 118}, 130405 (2017).

\bibitem{behnia} K. Behnia and A. Kapitulnik, A lower bound to the thermal diffusivity
of insulators. {\it J. Phys.: Condens. Matt.} {\bf 31}, 405702 (2019).

\bibitem{zaanen1} J. Zaanen, Y. Liu, Y. W. Sun and K. Schalm, Holographic duality in condensed matter physics (Cambridge University Press, 2015).

\bibitem{behnia1} Y. Machida, N. Matsumoto, T. Isono and K. Behnia, Phonon hydrodynamics and ultrahigh–roomtemperature
thermal conductivity in thin graphite. {\it Science}, {\bf 367}, 309 (2020).

\bibitem{hartnoll1} C. H. Mousatov and S. A. Hartnoll, On the Planckian bound for heat diffusion in insulators. {\it Nat. Phys.} {\bf 16}, 579-584 (2020).

\bibitem{sciadv} K. Trachenko and V. V. Brazhkin, Minimal quantum viscosity from fundamental physical constants. {\it Science Adv.} {\bf 6}, eaba3747 (2020).

\bibitem{landau} L. D. Landau and E. M. Lifshitz, Statistical Physics
(Oxford: Pergamon, 1969).

\bibitem{phillips} J. C. Phillips, Ionicity of the chemical bond in crystals. {\it Rev. Mod. Phys.} {\bf 42}, 317-356 (1970).

\bibitem{zaccone} B. Cui, A. Zaccone and D. Rodney, Nonaffine lattice dynamics with the Ewald method reveals strongly nonaffine elasticity of $\alpha$-quartz.
{\it J. Chem. Phys.} {\bf 151}, 224509 (2019).

\bibitem{diamond} V. V. Brazhkin, A. G. Lyapin and R. J. Hemley, {\it Phil. Mag.} {\bf 82}, 231-253 (2002).

\bibitem{diamond1} V. V. Brazhkin and V. L. Solozhenko, Myths about new ultrahard phases: Why materials that are significantly superior to diamond in elastic moduli and hardness are impossible. {\it J. Appl. Phys.} {\bf 125}, 130901 (2019).


\bibitem{handbook} CRC Handbook of Chemistry and Physics (ed. D. R. Lide, CRC Press, 2004).

\bibitem{handbook1} I. N. Frantsevich, F. F. Voronov and S. A. Bakuta, {\it Elastic constants and elastic moduli of metals and non-metals} (Kyiv, Naukova Dumka, 1982).

\bibitem{handbook2} Properties of Elements (Ed. M. E. Drits, Moscow Metallurgy, 1997).

\bibitem{correlation} B. Ratner, {\it Statistical and machine-learning data mining} (CRC Press, Taylor and Francis, 2011).

\bibitem{metals} T. Iida and R. I. L. Guthrie, {\it The Physical Properties of Liquid Metals} (Oxford University Press, 1988).

\bibitem{f2} V. V. Brazhkin and K. Trachenko, What separates a liquid from a gas? {\it Physics Today} 65(11), 68 (2012).

\bibitem{frenkel} J. Frenkel, {\it Kinetic Theory of Liquids} (Oxford University Press, New York, 1947).

\bibitem{boon} J. P. Boon and S. Yip, {\it Molecular Hydrodynamics} (New York: Dover, 1980).

\bibitem{f1} V. V. Brazhkin et al, ``Liquid-gas'' transition in the supercritical region: fundamental changes in particle dynamics. {\it Phys. Rev. Lett.} {\bf 111}, 145901 (2013).

\bibitem{f3} K. Trachenko and V. V. Brazhkin, Collective modes and thermodynamics of the liquids state. {\it Rep. Prog. Phys.} {\bf 79}, 016502 (2016).

\bibitem{gitis} M. B. Gitis and I. G. Mikhailov, On calculation of the speed of sound in liquid metals. {\it Acoustical Journal} {\bf 13}, 556-561 (1967) (in Russian).

\bibitem{vadim1} V. V. Brazhkin, Interparticle interaction in condensed media: some elements are ``more equal than others''. {\it Phys. Uspekhi} {\bf 52}, 369-376 (2009).

\bibitem{silvera} R. P. Dias and I. F. Silvera, Observation of the Wigner-Huntington transition to metallic hydrogen. {\it Science} {\bf 355}, 715-718 (2017).

\bibitem{loubeyre} P. Loubeyre, F. Occelli, and P. Dumas, Synchrotron infrared spectroscopic evidence of the probable transition to metal hydrogen. {\it Nature} {\bf 577}, 631-635 (2020).

\bibitem{hydrogen} J. M. McMahon, M. A. Morales, C. Pierleoni, and D. M. Ceperley, The properties of hydrogen and helium under extreme conditions. {\it Rev. Mod. Phys.} {\bf 84}, 1607-1653 (2012).

\bibitem{i41amd} K. Nagao, H. Nagara, and S. Matsubara, Structures of hydorgen at megabar pressures. {\it Phys. Rev. B} {\bf 56}, 2295-2298 (1997).

\bibitem{pickard_h_natphys} C. J. Pickard and R. J. Needs, Structure of phase III of solid hydrogen. {\it Nat. Phys.} {\bf 3}, 473 (2007).

\bibitem{azadi_metal} S. Azadi, B. Monserrat, W. M. C. Foulkes, and R. J. Needs, Dissociation of High-Pressure Solid Molecular Hydrogen: A Quantum Monte Carlo and Anharmonic Vibrational Study. {\it Phys. Rev. Lett.} {\bf 112}, 165501 (2014).

\bibitem{morales_metal} J. McMinis, R. C. Clay, D. Lee, and M. A. Morales, Molecular to Atomic Phase Transition in Hydrogen under High Pressure. {\it Phys. Rev. Lett.} {\bf 114}, 105305 (2015).

\bibitem{vadim2} V. V. Brazhkin and A. G. Lyapin, The inversion of relative shear rigidity in different material classes at megabar pressures. {\it J. Phys.: Condens. Matter} {\bf 14}, 10861-10867 (2002).

\bibitem{hadronic} P. Bedaque and A. W. Steiner, Sound Velocity Bound and Neutron Stars. {\it Phys. Rev. Lett.} {\bf 114}, 031103 (2015).

\bibitem{castep} S. J. Clark, M. D. Segall, C. J. Pickard, P. J. Hasnip, M. I. J. Probert, K. Refson, and M. C. Payne, First principles methods using CASTEP {\it Z. Kristallogr.} {\bf 220}, 567-570 (2005).

\bibitem{pbe} J. P. Perdew, K. Burke, and M. Ernzerhof, Generalized gradient approximation made simple, {\it Phys. Rev. Lett.} {\bf 77}, 3865-3868 (1996).

\bibitem{fd_martin} K. Kunc and R. M. Martin, Ab Initio Force Constants of GaAs: A New Approach to Calculation of Phonons and Dielectric Properties. {\it Phys. Rev. Lett.} {\bf 48}, 406-409 (1978).

\bibitem{nondiagonal} J. H. Lloyd-Williams and B. Monserrat, Lattice dynamics and electron-phonon coupling calculations using nondiagonal supercells. {\it Phys. Rev. B} {\bf 92}, 184301 (2015).

\end{thebibliography}
\end{document}